\documentclass[a4paper]{article}

\usepackage{amsmath,graphicx,epsfig,epstopdf,booktabs,mathtools}

\DeclareMathOperator*{\argmax}{arg\,max}
\DeclareMathOperator*{\isempty}{isempty}

\title{Deductive Refinement of Species Labelling in \\ Weakly Labelled Birdsong Recordings}

 \author{\em Veronica Morfi, Dan Stowell\\ \\ Machine Listening Lab, Centre for Digital Music, \\Queen Mary University of London, UK \\
  {\small \tt g.v.morfi@qmul.ac.uk, dan.stowell@qmul.ac.uk}
}
\begin{document}

  \maketitle
  \begin{abstract}
Many approaches have been used in bird species classification from their sound in order to provide labels for the whole of a recording. However, a more precise classification of each bird vocalization would be of great importance to the use and management of sound archives and bird monitoring. In this work, we introduce a technique that using a two step process can first automatically detect all bird vocalizations and then, with the use of `weakly' labelled recordings, classify them. Evaluations of our proposed method show that it achieves a correct classification of 61\% when used in a synthetic dataset, and up to 89\% when the synthetic dataset only consists of vocalizations larger than 1000 pixels.
  \end{abstract}

\section{Introduction}
\label{sec:intro}
The potential applications of automatic species detection and classification of birds from their sounds are many (e.g. ecology, archival). However, automated species identification is a challenging task due to the complexity of bird song, the noise present in most habitats, and the simultaneous song that occurs in many bird communities \cite{Luther:08} \cite{Luther:09}. Many authors have proposed methods for bird species classification (See \cite{Stowell:14} for a survey). However, more work is needed to address the problem of identifying all species and the exact times of their vocalizations in noisy recordings, containing multiple birds. Moreover, these tasks need to be achieved with minimal manual intervention, in particular without manual segmentation of recordings into birdsong syllables. Some of the early studies used small datasets, often noise-free and/or manually segmented and with a small number of species. More recent studies have fewer limitations, and introduce useful methods customised to the task \cite{Lakshminarayanan:09}  \cite{Damoulas:10} \cite{Briggs:12} \cite{Lee:2008}. However, these methods are only used for labelling the recordings (identifying the species present) and are not sufficient for detecting the exact times of the vocalizations. 

Techniques for automatic detection of audio events have, also, been of interest to many authors. Some of the techniques implemented so far are: a Markov-model based clustering for concept detection  \cite{Lee:10}, an SVM based method \cite{Lu:10}, a clustering and vector quantization to generate bag-of-audio-words representation to characterize audio and detect events \cite{Pancoast:12}, modeling classes as Gaussian component histograms on soundtracks of videos to identify types of videos \cite{Lee:10b}, and a speech recognition framework using HMMs for event detection \cite{Zhuang:08}.

In this work, our aim is to implement a two step process that using `weakly' labelled birdsong recordings can automatically detect each bird vocalization and then, classify it to one of these weak labels. By `weakly' labelled we refer to recordings that are annotated with which bird species are active, but have no information about which individual vocalizations are produced by which species. In order to implement our method, we propose a segmentation-detection process inspired by previously proposed ones (\cite{Fodor:13}, \cite{Lasseck:15} and \cite{Potamitis:15}), followed by a classification process based on finding the best visually similar match of a segment throughout the whole dataset and deductively refining its possible labels.

In the rest of the paper, Section \ref{sec:method} presents our proposed two step process. The evaluation follows in Section \ref{sec:evaluation} with the necessary discussions and conclusions in Section \ref{sec:conclusions}.

\section{Proposed Method}
\label{sec:method}

In this work, we are interested in extracting bird vocalizations in a fully unsupervised way and creating an algorithm that classifies them by using `weakly' labelled datasets. While manual annotation of the dataset is required to acquire the `weakly' labelled dataset, precise vocalization annotation is a much more time consuming process which requires expert knowledge. Additionally, there are already quite a few public datasets already labelled with the species present in each recording and a lot of methods that have been implemented in order to achieve a semi-automatic recording labelling \cite{Lakshminarayanan:09}  \cite{Damoulas:10} \cite{Briggs:12} \cite{Lee:2008}. To achieve our goal, we implement a two step process, first a segmentation-detection algorithm that detects all vocalizations, followed by a classification method that labels the segments in question.

For the segmentation-detection process, we employ the event detection para-digm used by G\'{a}bor Fodor \cite{Fodor:13}, Lasseck \cite{Lasseck:15} and Potamitis \cite{Potamitis:15}. This process is used in order to detect the specific coordinates of the bird vocalizations taking place in a spectrogram, disregarding any noise. All recordings labelled as having at least one bird species present are used in the extraction of segments.

Following this segmentation-detection process, a classification algorithm is also implemented. In our approach, `weakly' labelled recordings are used. Thus, each recording is labelled with the bird species it contains but we have no specific information as to when (time) and where (frequency) each vocalization takes place. What we attempt to do is classify each segment to one (or more) of the labels in the recording. During classification, each segment in need of labelling is matched via scikit-image's template matching function, \textit{match\_template}, to different recordings in order to obtain the best match and the most likely label. The results produced by \textit{match\_template} are processed by our classification algorithm in order to obtain the correct label. More detailed explanation on both processes (i.e. segmentation-detection and classification) can be found in the next subsections.

\subsection{Segmentation-Detection} 
\label{ssec:detection}
The unsupervised extraction of vocalization segments is of great importance to our classification task. Hence, we combined and refined the three already implemented segmentations as presented by G\'{a}bor Fodor in \cite{Fodor:13}, Lasseck in \cite{Lasseck:15} and Potamitis in \cite{Potamitis:15}, in order to create a segmentation process that best fits our automatic transcription task. All three methods are very closely related but have some differences. For all of them there are recordings where one will produce better results than the other two, in the sense of including segments that the other two failed to find or excluding segments produced by noise that the other two detected as vocalizations. In general, the methods proposed by Fodor and Potamitis will produce more segments than the one proposed by Lasseck. However, the Lasseck method is very effective in handling noise. It includes a spectral enhancement stage that reduces the number of segments produced by noise and discards acoustic events that fill the whole recording (e.g. Cicada songs, rain). For our purposes, a method that is robust to noise and does not generate noise segments is of great importance. Hence, we implement a close variation and refinement of these segmentations that is most similar to the Lasseck segmentation.

The following steps are performed for the spectrogram of each recording:
	\begin{itemize}
	\item normalize spectrogram to 1.0
	\item remove 4 lowest and 24 highest spectrogram rows
	\item get binary image via Median Clipping per frequency and time frame. We set pixel to 1 if its value is above 3 times the median of its corresponding row and column, otherwise it is set to 0
	\item apply closing in order to fill any small holes (pixels missing) in a vocalization
	\item remove small objects with size less than 10 pixels
	\item apply dilation and median filtering
	\item remove small objects with size less than 100 pixels
	\item re-apply dilation
	\item define all connected pixels as a segment ($seg_i$)
	\item find each segment's size and position with a small area of 5 pixels added to each direction
	\end{itemize}

%\begin{center}
%		\includegraphics [width=0.8\linewidth,clip, trim=0mm 0mm 0mm 0mm]  {segments.png}
%	\end{center}

\subsection{Classification}
\label{ssec:classification}
In our method, `weakly' labelled recordings are used. Hence, the species present are the labels of that recording ($labels\_rec$), however, we have no further information as to the specific vocalizations. For each recording, the segments that derive from the segmentation-detection process ($seg_i$) are considered to be the product of vocalizations from the bird species included in the weak labels. For each segment, we create a list of possible labels ($labels\_seg_i$), initialized to the weak labels of the recording that contains the segment. The $labels\_seg_i$ list will later on be shortened to either one or multiple labels by the classification process. 

During classification, scikit-image's \textit{match\_template} function is used to compute the similarity between a segment and a target image. The highest matching probability is computed by using normalized cross-correlation between the two. Due to the number of recordings and segments detected in each of them, this process is very time consuming. However, similar bird sounds should appear in similar frequencies, hence we can reduce the computational load by only applying \textit{match\_template} to a smaller range of frequencies (5 pixels below and above the segment frequencies). 

Furthermore, since the weak labels of a recording, and a segment, are already known, we only need to search recordings that contain at least one of them. In order to best utilize the information provided by the weak labels, we first find all the recordings that contain at least one of the $labels\_rec$ and then group them depending on the number of the weak labels they contain (1 of the weak labels, 2 of the weak labels, etc.). These recordings may contain other labels, too.

More specifically, during the classification process we obtain the groups of recordings $recs(c)$, where $c$ denotes the different label combinations produced by $labels\_rec$. The recordings in $recs(c)$ have label(s) $c$ present in their weak labels. For each recording, we first create a group of recordings that only contain one of the possible labels, followed by recordings that contain two, three, etc. Function \textit{match\_template} will first try to find a match in the recordings that only contain one of the possible labels. A match is found when the similarity rate returned by \textit{match\_template} is 0.4 or greater (\textit{match\_template}'s results can range from -1.0 to 1.0). All the different matches (similarity $\geq$ 0.4) found in these recordings will be, later, summed and the label with the highest sum will be the one most likely for the segment, hence the segment list of possible labels will get reduced to that single label. If no match is found in the group containing one possible label then the group of recordings containing two possible labels is checked, followed by the next group (3 possible labels) if no match is found, etc. When a match is found in a recording that contains more than one of the possible labels, the segment's possible label list is reduced to all the possible matches. 

The above classification process was implemented in three different procedures, namely the First-Pass, 1st variation and 2nd variation of the classification. All three are applied to the recordings in order, as explained in the following subsections and illustrated in Figure \ref{fig:fig1}.

\subsubsection{First-Pass}
\label{original}

\noindent The first pass of the classification follows the aforementioned matching process. Each segment in need of a label will search through the list of recordings $recs(c)$ that first contain only one of the possible labels ($|c|=1$). If a match is found between these, the segment will get the common label between its possible labels and the labels of the recording with the best match $C$ (greatest sum of all matches for each label). If no match is found, the algorithm will proceed with the list of recordings that has two of the possible labels ($|c|=2$). If a match is found within them and the segment, then the segment's possible labels list will be reduced to the common labels between the initial possible label list and the matching recording's labels: $labels\_seg_i = \cap(C, labels\_seg_i)$. The algorithm will continue with recordings which containing three, four, etc. of the labels, until a match is found or until there are no more recordings. If no match is found the Match Not Found ($MNF$) label is assigned to the segment. Segments with the $MNF$ label and segments that have more than one possible label are classified as Unknown in our evaluation results (Section \ref{sec:evaluation}), even if the correct segment label is between the multiple possible labels. Algorithm 1 describes this classification procedure.

\begin{center}
\begin{tabular}{l}
 \hline\noalign{\smallskip} 
  \hline\noalign{\smallskip} 
\textbf{Algorithm 1:} Classification Process (First-Pass)\\
 \hline\noalign{\smallskip}   
 \hline\noalign{\smallskip} 
\textbf{for} each segment $i=1:total\_segments$: \\
\hspace{0.3cm} $labels\_seg_i = labels\_rec$ \\
\hspace{0.3cm} \textbf{for} $j = 1:length(labels\_rec)$: \\
\hspace{0.6cm} \textbf{for} each combination $c$ of $j$ labels:\\
\hspace{0.9cm} $recs(c)=$ recordings that contain only the labels in $c$ \\
\hspace{0.9cm} \textbf{if} $match\_template(seg_i, recs(c)) \geq 0.4$: \\
\hspace{1.2cm} $match_{i}(c) = \sum{match\_template(seg_i, recs(c))}$\\
\hspace{0.9cm} end \textbf{if}\\
\hspace{0.6cm} end \textbf{for}\\
\hspace{0.6cm} $C = \argmax(match_{i}(c))$ \\
\hspace{0.6cm} \textbf{if} $\isempty(C)$: \\
\hspace{0.9cm} \textbf{continue for}\\
\hspace{0.6cm} end \textbf{if}\\
\hspace{0.3cm} end \textbf{for}\\
\hspace{0.3cm} \textbf{if} $\isempty(C)$: \\
\hspace{0.6cm} $labels\_seg_i = MNF$\\
\hspace{0.3cm} \textbf{else}:\\
\hspace{0.6cm} $labels\_seg_i = \cap(C, labels\_seg_i)$\\
\hspace{0.3cm} end \textbf{if}\\
end \textbf{for}\\
 \hline   
 \label{tab:original}
\end{tabular}
\end{center}

\subsubsection{1st Variation}
\label{1st}

\noindent The 1st variation of the process derived from the need to solve the relatively large amount of $MNF$ labelled segments produced through the first-pass of the classification. Since we use only weakly labelled datasets, all the labels of a recording must be assigned to at least one segment. A trivial solution of reducing the $MNF$ segments is: when there are $MNF$ segments and labels with no segments in a recording ($c_{un}$), we assign the unallocated labels to all the $MNF$ segments. This will solve the issue of unallocated labels in a recording but will not completely eliminate the Unknown segments, since more than one label may be assigned to a single segment. Algorithm 2 describes the 1st variation.

\begin{center}
\begin{tabular}{l}
 \hline\noalign{\smallskip} 
  \hline\noalign{\smallskip} 
\textbf{Algorithm 2:} Classification Process (1st Variation)\\
 \hline\noalign{\smallskip}   
 \hline\noalign{\smallskip} 
\textbf{if} unallocated label(s) $c_{un}$ \textbf{and} any $labels\_seg_i=MNF$:\\
\hspace{0.3cm} $labels\_seg_i =$ $c_{un}$, $\forall MNF$ segments \\
end \textbf{if}\\
\hline   
 \label{tab:1stvar}
\end{tabular}
\end{center}

\subsubsection{2nd Variation}
\label{2nd}

\noindent After reducing the $MNF$ segments in the whole dataset, there may still be labels unallocated in some recordings. Hence, the 2nd variation derived from the need for all labels of a recording to get assigned to at least one segment. This approach is designed for recordings that never had any $MNF$ segments to begin with. More specifically, if there are labels in recordings that are not assigned to any segments but all their segments have a label, then there are some labels that are assigned to more than one segments. It is possible that more than one segment may have the same label, but when a label is unallocated then it is assumed that one of those segments matched to the same label is wrongly classified. A match for the unallocated labels will be searched for within these segments. We search for the best match for any unallocated label, if that exists, between the multiple segments of the rest of the labels. If a match is found, the label of the segment it derives from is changed to the unallocated label. Algorithm 3 explains the 2nd variation.

\begin{center}
\begin{tabular}{l}
 \hline\noalign{\smallskip} 
  \hline\noalign{\smallskip} 
\textbf{Algorithm 3:} Classification Process (2nd Variation)\\
 \hline\noalign{\smallskip}   
 \hline\noalign{\smallskip} 
\textbf{if} unallocated label(s) $c_{un}$ \textbf{and} $|$segments with label $c| \geq 2$, $\forall c$  :\\
\hspace{0.3cm} $same(c) =$ all segments labelled $c$ \\
\hspace{0.3cm} $match(c_{un}) = match\_template(same(c),recs(c_{un}))$\\
\hspace{0.3cm} Find segment $seg_i$ with $\max(match(c_{un}))$  \\
\hspace{0.3cm} $labels\_seg_i = c_{un}$\\
end \textbf{if}\\
 \hline   
 \label{tab:2ndvar}
\end{tabular}
\end{center}
\vspace{-0.5cm}

\begin{figure}[!htb]
\centering
\includegraphics[width=8cm, height= 18cm]{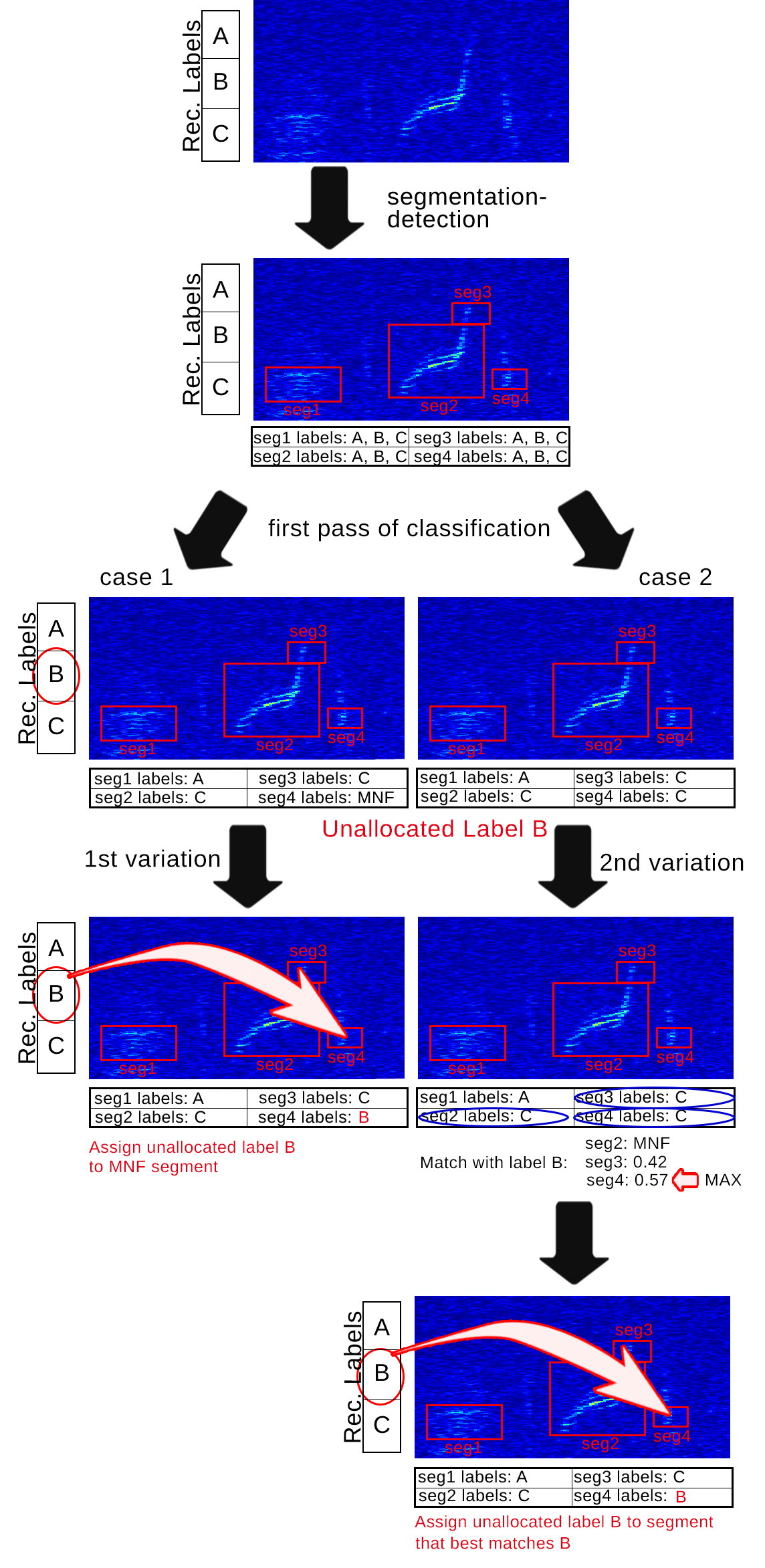}
\caption{Example of the proposed two step process. Case 1 describes what happens when there is an unallocated label and a segment with $MNF$ label. Case 2 describes what happens when there is an unallocated label and multiple segments have one of the other labels.}
\label{fig:fig1}
\end{figure}

Figure \ref{fig:fig1} provides a visual example of all the above classification steps. Case 1 depicts what happens when there is an unallocated label (label B) and one of the segments has $MNF$ label (segment 4), hence the 1st variation is used. In this case, the unallocated label will be assigned to segment 4. In case 2, all segments have labels assigned to them, but still label B is not assigned to any of them. Hence, the 2nd variation will search for the best match with label B within the segments that have the same label (segments 2, 3 and 4). Segment 4 has the max match of 0.57, thus label B will be assigned to it.

\section{Evaluation}
\label{sec:evaluation}
To our knowledge, there is currently no public dataset with strong time-frequency labelling of each bird vocalization. Thus, in order to evaluate our proposed method and its variations, a synthetic dataset was created. Using the audio dataset provided during the Neural Information Processing Scaled for Bioacoustics (NIPS4B) bird song competition of 2013\footnote{http://sabiod.univ-tln.fr/nips4b/challenge1.html}, we created a synthetic dataset of 50 recordings. Out of the 87 labels of the NIPS4B dataset, 37 have recordings that are labelled with only one species. Those recordings, and hence labels, are the ones used for our synthetic dataset. Each synthetic recording is 5 seconds long and it consists of a noise recording, with no labels, randomly picked from the NIPS4B dataset. Each synthetic recording is allocated 2 to 5 randomly picked labels out of the above mentioned 37 labels. A source recording is randomly picked for each of the labels and from that recording one segment is placed in the synthetic recording. Thus, each synthetic recording contains 2 to 5 segments. The resulting dataset consists of 50 recordings, with a total of 152 segments. For this evaluation, there is no overlapping of vocalizations simultaneously in both time and frequency in any of the synthetic recordings. We used the remainder of the original NIPS4B dataset in order to search for the segment matches during the classification process.

\begin{table}[!htb]
\caption{Classification Results for Synthetic Dataset}
\begin{center}
\begin{tabular}{l c c c}
\toprule
\textbf{} & \textbf{Correct}    & \textbf{Wrong} & \textbf{Unknown} \\
\midrule 
\textbf{Chance} & 33.3\%      & 66.7\%    &  --- \\
\textbf{First-Pass} & 54\%      & 29.5\%    & 16.5\% \\
\textbf{1st Variation} & 59\%      & 32\%    & 9\% \\
\textbf{2nd Variation} & 61\%      & 30\%    & 9\% \\
 \bottomrule
\end{tabular}
\label{table:results}
\end{center}
\end{table}

In Table \ref{table:results}, the results of the segment classification using all three methods are depicted. The original classifier (First-Pass) produces a correct classification of 54\% and 16.5\% of Unknown segments, the latest one includes segments that are either not matched to anything ($MNF$ label) or have more than one labels. After we apply the 1st variation of the algorithm, the percentage of Unknown segments is reduced to 9\%, while both the correctly and wrongly classified segments are increased. Finally, once we apply the 2nd variation, we have a slight increase to the number of correct classifications, namely 2\%, which leads to the total result of 61\% correctly classified segments. 

For this dataset, most of the misclassifications happen due to the fact that the segmentation-detection process produces a lot of segments with size less than 1000 pixels and those segments usually contain very simple vocalizations, and in many cases, fragments of vocalizations, that can be matched to multiple labels easily. In the case of the segments being part of vocalizations the classification results can be verified through a process of inverse matching. More explicitly, checking the recording where a match is found to see if it was matched to a single segment or a part of a bigger segment, by checking the area around where the match was found. If the segment is matched to part of a bigger vocalization then it must have the remaining of the vocalization at a close by area in order for it to be considered a correct classification. However, the above problem cannot be solved in the synthetic dataset case, because the segments are chosen at random, so they are not placed together with the rest of the vocalization. These segments are considered `out of context'. 

In order to evaluate the classification methods when the `out of context' problem does not occur, we created another, smaller, synthetic dataset of 13 recordings, where all the segments had to be over 1000 pixels. In this dataset, each recording contains 2 to 4 labels, hence 2 to 4 segments and there is a total of 36 segments. The results produced by the different classifications are shown in Table \ref{table:results1000}.

\begin{table}[!htb]
\caption{Classification Results for Segments $\geq$ 1000 pixels}
\begin{center}
\begin{tabular}{l c c c}
\toprule
\textbf{} & \textbf{Correct}    & \textbf{Wrong} & \textbf{Unknown} \\
 \midrule
\textbf{Chance} & 33.3\%      & 66.7\%    &  --- \\
\textbf{First-Pass} & 53\%      & 8\%    & 39\% \\
\textbf{1st Variation} & 83\%      & 11.5\%    & 5.5\% \\
\textbf{2nd Variation} & 89\%      & 5.5\%    & 5.5\% \\
 \bottomrule
\end{tabular}
\label{table:results1000}
\end{center}
\end{table}

A great increase in the number of Unknown segments can be noticed in the results produced by the first-pass of the classification process compared to the ones on Table \ref{table:results}. This is due to the fact that only big segments ($\geq$ 1000 pixels) are used. Some birds in the NIPS4B dataset produce very complex and unique, even between their own species, vocalizations that cannot be matched to anything else, at least visually. Hence, when we only choose segments with size greater than 1000 pixels the chance of selecting more of those complex segments increases compared to the case of the previous dataset. Therefore, there is a 39\% of Unknown segments, most of them being segments labelled as $MNF$. However, once the 1st variation is applied we notice a great increase in correct classification, namely an increase of 30\%, leading to a result of 83\% correct classification. Finally, after the 2nd variation is applied, once again, there is a slight increase (6\%) of correct classification that results to 89\% correct classified segments.

\section{Conclusions}
\label{sec:conclusions}
Taking advantage of the good bird species classification results produced by image segmentation and event detection methods, we proposed a two step process that can be applied to `weakly' labelled recordings. Our approach implements a fully automatic way of extracting the bird vocalizations in each recording using its corresponding spectrogram. Additionally, by utilizing the information provided by the weak labels of a recording, we are able to reduce the possible labels of each detected vocalization. According to the assessment of correct classification, in our synthetic dataset, our two step process achieves up to 61\% successful classification per vocalization. Moreover, when used in a synthetic dataset compiled by segments with size greater than 1000 pixels, it attains up to 89\% correct classification rate.

  \newpage
%  \eightpt
  \bibliographystyle{IEEEtran}

  \bibliography{interspeech2016}

%  \begin{thebibliography}{9}
%    \bibitem[1]{Davis80-COP}
%      S.\ B.\ Davis and P.\ Mermelstein,
%      ``Comparison of parametric representation for monosyllabic word recognition in continuously spoken sentences,''
%      \textit{IEEE Transactions on Acoustics, Speech and Signal Processing}, vol.~28, no.~4, pp.~357--366, 1980.
%    \bibitem[2]{Rabiner89-ATO}
%      L.\ R.\ Rabiner,
%      ``A tutorial on hidden Markov models and selected applications in speech recognition,''
%      \textit{Proceedings of the IEEE}, vol.~77, no.~2, pp.~257-286, 1989.
%    \bibitem[3]{Hastie09-TEO}
%      T.\ Hastie, R.\ Tibshirani, and J.\ Friedman,
%      \textit{The Elements of Statistical Learning -- Data Mining, Inference, and Prediction}.
%      New York: Springer, 2009.
%    \bibitem[4]{YourName16-XXX}
%      F.\ Lastname1, F.\ Lastname2, and F.\ Lastname3,
%      ``Title of your INTERSPEECH 2016 publication,''
%      in \textit{Interspeech 2016 -- 16\textsuperscript{th} Annual Conference of the International Speech Communication Association, September 8–12, San Francisco, California, USA, Proceedings, Proceedings}, 2016, pp.~100--104.
%  \end{thebibliography}

\end{document}